\documentclass[conference]{IEEEtran}
\IEEEoverridecommandlockouts

\usepackage{cite}
\usepackage{spconf,amsmath,amssymb,amsfonts}
\usepackage{graphicx}
\usepackage{textcomp}
\usepackage{xcolor}
\usepackage{algorithm}
\usepackage{algpseudocode}
\usepackage{tikz}
\usepackage{xfrac}
\usepackage{caption}
\usepackage{subcaption}

\usetikzlibrary{shapes.geometric, arrows, chains}

\tikzstyle{algoblock} = [rectangle, rounded corners, 
minimum width=3cm, 
minimum height=1cm,
text centered, 
draw=black]

\tikzstyle{sum} = [draw, circle, minimum size=.5cm, node distance=1.75cm]

\tikzstyle{rect} = [draw, rectangle, minimum height = 4em, text width = 6em, text centered]
\tikzstyle{arrow} = [thick,->,>=stealth]

\DeclareMathOperator*{\argmin}{arg\,min}
\def\BibTeX{{\rm B\kern-.05em{\sc i\kern-.025em b}\kern-.08em
    T\kern-.1667em\lower.7ex\hbox{E}\kern-.125emX}}

\begin{document}

\title{Dynamic DH-MBIR for Phase-Error Estimation from Streaming Digital-Holography Data}

\name{
\begin{tabular}{@{}c@{}}
Ali G. Sheikh$^{1}$ \qquad 
Casey J. Pellizzari$^{2}$ \qquad 
Sherman J. Kisner$^{3}$ \\[10pt]
Gregery T. Buzzard$^{1}$ \qquad 
Charles A. Bouman$^{1}$ 
\end{tabular}}

\address{$^{1}$ Purdue University, West Lafayette, IN  \\
     $^{2}$ United States Air Force Academy, Colorado Springs, CO \\
     $^{3}$ High Performance Imaging LLC, West Lafayette, IN
\thanks{This research was supported by the United States Air Force and the Air Force Research Laboratory. GTB was partially supported by NSF CCF-1763896.  CAB was partially supported by the Showalter Trust. 
Approved for public release: distribution unlimited.
The views expressed in this article, book, or presentation are those of the author and do not necessarily reflect the official policy or position of the United States Air Force Academy, the Air Force, the Department of Defense, or the U.S. Government.
}
}

\maketitle

\begin{abstract}

Directed energy applications require the estimation of digital-holographic (DH) phase errors due to atmospheric turbulence in order to accurately focus the outgoing beam.
These phase error estimates must be computed with very low latency to keep pace with changing atmospheric parameters, which requires that phase errors be estimated in a single shot of DH data. 
The digital holography model-based iterative reconstruction (DH-MBIR) algorithm is capable of accurately estimating phase errors in a single shot using the expectation-maximization (EM) algorithm.
However, existing implementations of DH-MBIR require hundreds of iterations, which is not practical for real-time applications.

In this paper, we present the Dynamic DH-MBIR (DDH-MBIR) algorithm for estimating isoplanatic phase errors from streaming single-shot data with extremely low latency. The Dynamic DH-MBIR algorithm reduces the computation and latency by orders of magnitude relative to conventional DH-MBIR, making real-time throughput and latency feasible in applications. Using simulated data that models frozen flow of atmospheric turbulence, we show that our algorithm can achieve a consistently high Strehl ratio with realistic simulation parameters using only 1 iteration per timestep.

\end{abstract}


\section{Introduction}

The goal of a directed energy system is to focus a beam of light on a distant object. 
In terrestrial applications, this can be difficult because atmospheric turbulence introduces random phase errors (or distortions) that tend to disperse the otherwise focused beam.
A solution to this problem is to pre-distort the phase of the outgoing beam.
This can be done by first performing wavefront sensing (WFS), in which the phase distortion of incoming light is measured.
The outgoing light can then be pre-distorted with the conjugate phase.
However, this approach requires that the WFS be performed with very low latency, so that the phase corrections can be used before the atmospheric distortion changes.

Traditionally, WFS is performed with a Shack-Hartmann wavefront sensor.
However, since the Shack-Hartmann method is based on hardware, it is limited in speed and is effective only with moderate turbulence strength \cite{barchers2002evaluation}.

Alternatively, digital holography (DH) sensors can be used to estimate the wavefront from direct measurements of the light.
DH sensors measure the amplitude and phase of incoming light by interfering the received light with a reference beam. 
The resulting complex-valued measurements can then be used to computationally estimate phase distortions by solving an inverse problem  \cite{spencer2016}.

The Image Sharpening (IS) algorithm is one approach to estimating phase distortions from DH data \cite{thurman2008}.
This method is optimization-based and models the phase distortion with a Zernike expansion. 
However, the measured reflection from coherent light contains speckle \cite{Goodman06}, which reduces the effectiveness of the IS algorithm.
A solution to this speckle problem is to take multiple shots of DH data in which the speckle is decorrelated, but this increases the latency of the WFS, which is not acceptable in directed energy systems.

The DH model-based iterative reconstruction (DH-MBIR) approach in \cite{pellizzari2017} uses a probabilistic model that accounts for speckle together with prior information or regularization.  
This approach represents a target image using the real-valued reflectance, which is smoother than the complex-valued reflection coefficients associated with speckle; hence reflectance is more amenable to regularization.  
This approach yields state-of-the-art estimates of images and pupil-plane phase errors using single-shot DH data.
However, the current version of DH-MBIR is designed to process a single shot of DH data, and it requires hundreds of iterations per shot, hence is not practical in a real directed energy system.

In this paper, we introduce the Dynamic DH-MBIR algorithm (DDH-MBIR) that can reconstruct a streaming sequence of single-shot DH data with only 1 iteration per shot.
This results in a low-latency, high-throughput algorithm for dynamic reconstruction of phase errors due to atmospheric turbulence.

The DDH-MBIR algorithm has two key advantages:
\begin{itemize}
    \item It uses temporal correlation in both $\phi$ and $r$ to dramatically reduce the number of iterations per time-step.
    \item It avoids poor quality solutions associated with local minima by incorporating the bootstrap step into the dynamic update process.
\end{itemize}

Our simulation results demonstrate that under realistic assumptions, the Dynamic DH-MBIR algorithm can achieve highly accurate, low latency phase error estimation in a streaming system using single-shot DH data in each frame.

\section{Forward Model}

A DH sensor allows us to measure the complex electromagnetic field reflected from an object using spatial-heterodyne interferometry. 
As in \cite{pellizzari2017}, we model the complex-valued measurements at time $n$ as 
\begin{equation} \label{eq:yn}
y_n=A_{\phi_n} g_n + w_n,
\end{equation}
where $y_n \in \mathbb{C}^M$ is the rasterized vector of complex DH measurements,
$g_n \in \mathbb{C}^M$ is the vector of unknown complex reflection coefficients from the illuminated object, 
$w_n \in \mathbb{C}^M$ is a vector of complex measurement noise,
and $A_{\phi_n} \in \mathbb{C}^{M\times M}$ is a linear transformation that is dependent on the unknown phase error, $\phi_n$.

If the wavelength of the light is small compared to the resolved pixel size on the object,
then the reflection coefficient, $g_n$, is accurately modeled as a circularly symmetric, complex Gaussian with unknown variance, $r_n$.
In this case, $r_n$ is real-valued reflectance with $r_n = E[|g_n|^2]$ and $g_n \sim CN ( 0, r_n )$.

We assume an isoplanatic propagation model in which the phase distortion, $\phi_n$, occurs close to the detector \cite{pellizzari2019a} (however, our proposed method also applies in the anisoplanatic case).
The isoplanatic linear forward model can be decomposed as
$$
A_{\phi_n} = D_a {\cal D} (e^{j{\phi_n}})F \, \Gamma
$$
where $D_a$ is a diagonal matrix that models the camera aperture,
${\cal D} (e^{j{\phi_n}})$ is a diagonal matrix of phase distortions,
$F$ is a normalized 2D discrete Fourier transform (DFT), 
and $\Gamma$ is a diagonal matrix of quadratic phase factors resulting from Fresnel propagation \cite{schmidt2010}.

\section{The DH-MBIR Algorithm}

Our goal is to estimate the unknown phase errors, $\phi_n$, from the measurements, $y_n$.
To do this, we will also need to estimate the object reflectance, $r_n$.
Importantly, the reflectance is typically a smoother quantity with higher
spatial correlation relative to the reflection coefficient,
$g_n$, hence is more amenable to accurate estimation.
Thus, our goal will be to compute at each time-step the joint-MAP estimate 
\begin{equation}   
\label{eq:MAP}
(\hat{r}_n,\hat{\phi}_n ) = \argmin_{r,\phi} \{ -\log p(y_n|r,\phi)-\log p(r) -\log p(\phi) \} 
\end{equation}
Direct minimization of~\eqref{eq:MAP} is not tractable due to the nonlinear relationship between $r_n$ and $y_n$.
However, the DH-MBIR algorithm solves this problem by using the expectation-maximization (EM) algorithm to solve for the MAP estimate using the iterative application of surrogate functions \cite{pellizzari2017}.

The DH-MBIR algorithm has the form

\smallskip
\centerline{\fbox{
\hspace*{-10pt}
\parbox{1.6in}{
\begin{algorithmic}
  \State $r \gets 0$; $\phi \gets 0$
  \While{not converged} \; \;
      $(r,\phi ) \leftarrow EM( r, \phi ; y )$
  \EndWhile 
    \end{algorithmic}
}}}
\smallskip

\noindent 
where the function $EM(r,\phi; y)$ computes one iteration of the EM algorithm.
We also note that the computation of the $EM()$ function is dominated by the application of two FFTs \cite{SridharKisner2020} and has roughly the same computational cost as one iteration of the IS algorithm.

However, the existing DH-MBIR algorithm is impractically slow for two reasons.
First, it requires hundreds of iterations of the $EM()$ function. 
Second, since the underlying function being minimized is non-convex, the algorithm tends to become trapped in local minimum.
This second problem has been largely solved using a bootstrapping technique proposed by Pellizzari \cite{pellizzari2017b} in which the reconstructed image $r$ is reset to a backprojected estimate every 200 iterations.

\section{Dynamic DH-MBIR Algorithm}

\begin{figure}
\centering
\includegraphics[width=0.5\textwidth]{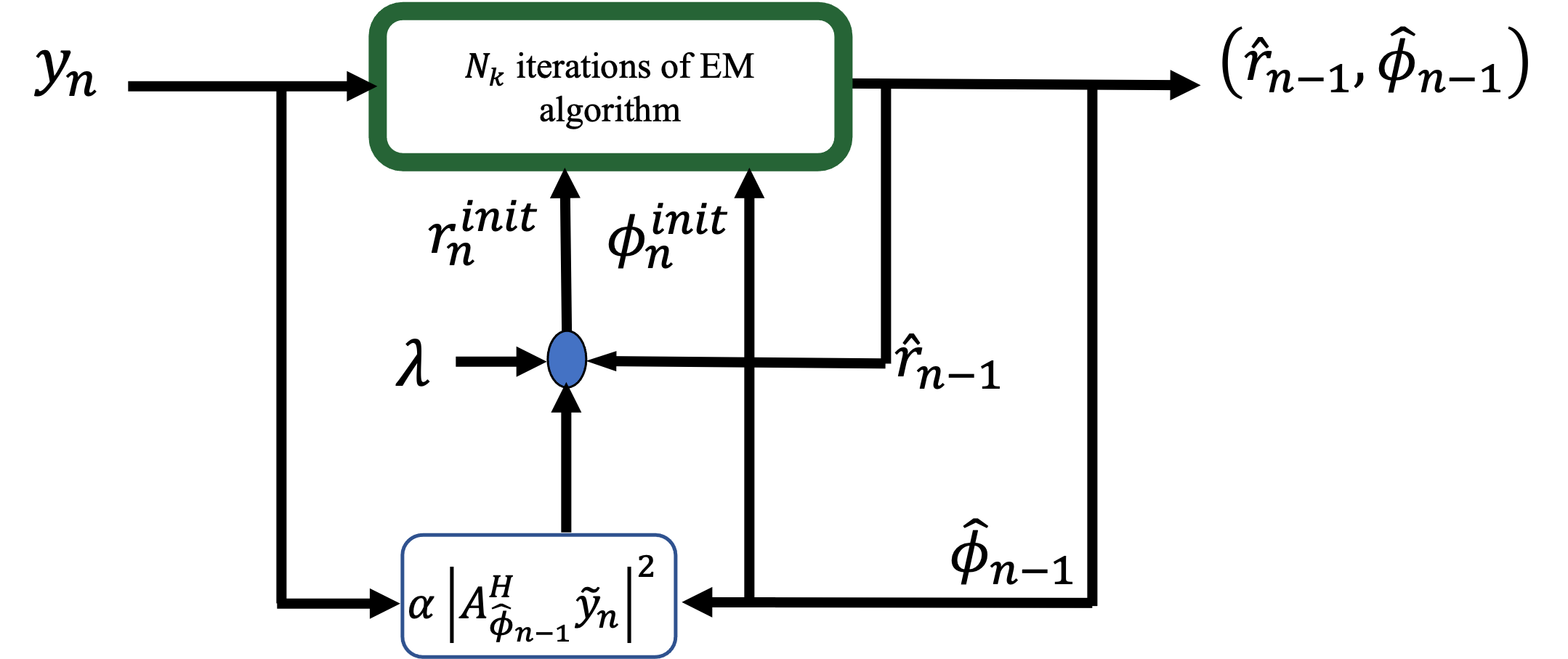}
\caption{Dynamic DH-MBIR}
\label{fig:dynamic-dhmbir-flowchart}
\end{figure}

In directed energy applications, $\phi_n$ must be estimated at a sufficiently high sampling rate, $f_s$, that the atmospheric phase distortion is nearly unchanged between samples.
For this, we propose the Dynamic DH-MBIR (DDH-MBIR) algorithm for estimating $r_n$ and $\phi_n$ from a streaming sequence of DH data, $y_n$. 
Figures~\ref{fig:Dynamic-DHMBIR} and~\ref{fig:dynamic-dhmbir-flowchart} show the pseudo-code and a flow diagram for DDH-MBIR.
With each iteration, the data is read, normalized, and then residual tip/tilt removed using least-squares regression.
Next, the value of $r_n$ is initialized, and $N_k$ iterations of the EM algorithm are performed.

\begin{figure}
\center{
\fbox{\parbox{2.5in}{
\begin{algorithmic}
\Function{DDH\_MBIR}{$\lambda, \alpha, N_k$}
  \State $n \gets 0$; $r \gets 0$; $\phi \gets 0$
  \While{Data is available}  
    \State $y \leftarrow \mbox{ReadData} (n)$
    \State $y \leftarrow \mbox{Normalize} ( y )$
    \State $\phi \leftarrow \mbox{RemoveTipTilt} (\phi )$
    \State $r \gets (1-\lambda ) r + \lambda \alpha \left| A_{\phi}^H y \right|^2$
    \For{$i=0$ to $N_k-1$}
      \State $(r,\phi) \gets EM( r , \phi ; y )$
    \EndFor
    \State WriteData($n,r,\phi$)
    \State $n \gets n +1$
  \EndWhile
\EndFunction
\end{algorithmic}
}}}
\caption{Pseudo-code for the Dynamic DH-MBIR algorithm. Each iteration initializes $r$ with a weighted combination of the previous estimate and a back-projection of the DH data.}
\label{fig:Dynamic-DHMBIR}
\end{figure}

The key innovation of the DDH-MBIR algorithm is the initialization of $r_n$ given by
\begin{equation}
r_n^{init} \gets (1-\lambda ) \underbrace{ {r}_{n-1} }_{1^{st}\rm\ term} 
    + \lambda \alpha \underbrace{ | A_{{\phi}_{n-1}}^H y_n |^2 }_{ 2^{nd}\rm\ term} \ .
\end{equation}
This initialization leverages the temporal correlation of both $\phi$ and $r$ to dramatically reduce the number of EM iterations per time step, $N_k$ (we show later that $N_k=1$ yields high quality estimates of the phase distortion on simulation data under practical conditions).  
This initialization is formed by a weighted sum of two terms where $\lambda$ and $\alpha$ are user selectable weights in the range $[0,1]$.
Since $r$ does not typically change substantially between samples, the first term, $r_{n-1}$, provides a good initial condition for the estimation of $r_n$.

The second term uses the magnitude squared of the back-projected data. This is typically how images are formed with DH data, and in particular, this is the term that is used in the bootstrapping algorithm of \cite{pellizzari2017b}.
Intuitively, this second term incorporates a small amount of continuous bootstrapping into the dynamic update loop, which keeps the DDH-MBIR output away from solutions associated with local minima.

Finally, the \mbox{RemoveTipTilt}() function removes the linear component of the phase corresponding to spatial shifts of the reconstructed image;
and the \mbox{Normalize}() function is defined as 
\begin{equation}
\text{Normalize}(y) = \sqrt{p} \frac{y - \mu}{\| y - \mu \|} \ ,
\end{equation}
where $\mu$ is the mean value of $y$ and $p$ is the number of entries in $y$.  This normalization allows algorithmic parameters to be chosen more robustly.


\section{Experimental Results}
In this section, we present estimation results from synthetic data using the Dynamic DH-MBIR algorithm.

\subsection{Data Simulation}

We generated data using the 1951 USAF resolution test chart for $r$ along with a frozen flow phase-error model as specified in \cite{srinath2015} for $\phi$.
We also removed the piston, tip, and tilt of the phase-errors because we are interested only in estimating the higher-order aberrations. 
At each timestep, the DH data, $y_n$, was generated as in \cite{pellizzari2017}.

Furthermore, we used an isoplanatic model, image and phase distortion array sizes of $N \times N$ where $N=256$, a sampling frequency of $f_s=10kHz$, a Greenwood frequency $f_g = 100Hz$, a turbulence strength of $D/r_0 = 10$ where $D$ is the diameter of the aperture and $r_0$ is Fried's parameter \cite{fried1990}, 
a wavelength of $1.064 \mu m$, and an SNR of $10dB$.

Using these parameters, we find the magnitude of the phase displacement between time samples by scaling Equation 4.15 in \cite{tyson_introduction_2000} to convert from m/s to pixels/sample.  This gives
$$
\|v\| = \frac{1}{0.43} \frac{f_g}{f_s} \frac{N}{D/r_0} \ .
$$
In our simulation we assumed a flow downward and to the right, which gives a vector displacement of  $v=(0.421,0.421)$ pixels per time step.

\subsection{Algorithm Parameters and Metrics}

In all our simulations, the DDH-MBIR parameters are $\alpha =0.025$, $\lambda = 0.45$,
which were found to be the best in a grid search over the range $\alpha \in [0, 0.1]$ and $\lambda \in [0.1]$.

We used peak Strehl ratio, $S$, to evaluate the quality of the phase-errors estimates, where 
$$
S = \frac{[|A_{\hat{\phi}_n-\phi_n}^H D_a |^{ 2}]_{\max}}{[|A_{0}^H D_a |^{ 2}]_{\max}} \ ,
$$ 
and $\hat{\phi}_n$ is the estimated phase, $A_{0}^H$ is back-propagation through a vacuum, and $[\cdot]_{\max}$ indicates the maximum value of the argument.

\subsection{Results}

\begin{figure}
\centerline{
\includegraphics[width=0.45\textwidth]{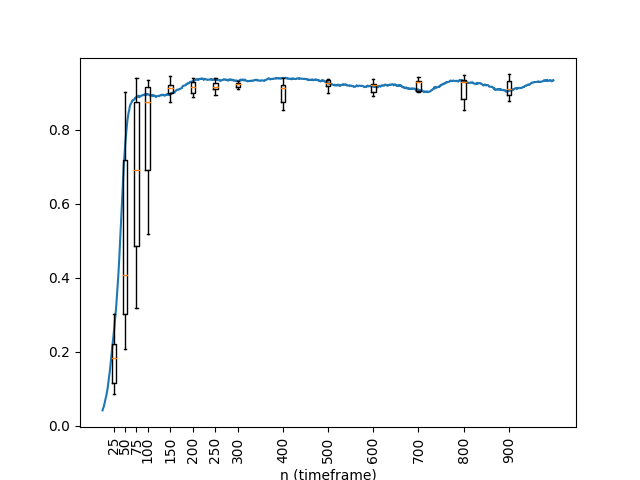}
}
\caption{Box plot of Strehl ratio for 10 simulations of $N_k=1$ iteration/timeframe of DDH-MBIR with the Strehl curve of a particular simulation overlaid in blue.}
\label{fig:N_k=1}
\end{figure}

\begin{figure}
    \centering
    \includegraphics[width=0.45\textwidth]{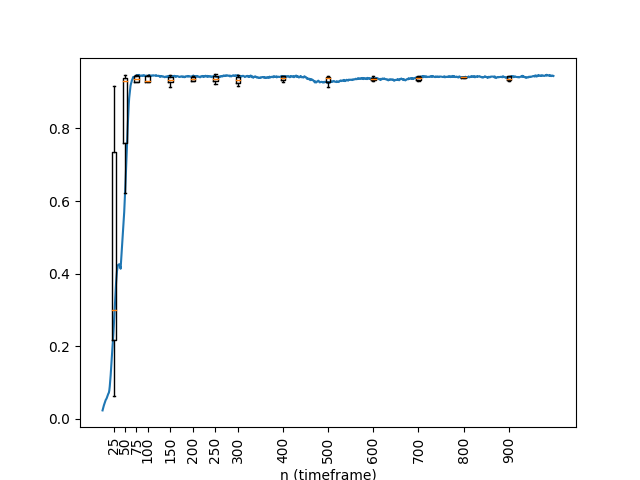}
    \begin{picture}(0,0)
\put(-185,20){\includegraphics[height=4.cm]{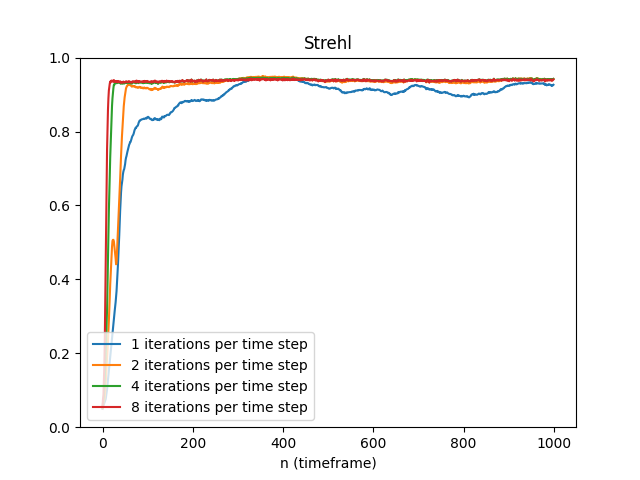}}
\end{picture}
\caption{Box plot of Strehl ratio for 10 simulations of $N_k=2$ iteration/timeframe of DDH-MBIR with the Strehl curve of a particular simulation overlaid in blue. Inset shows Strehl ratios as a function of time frame for $N_k = 1, 2, 4, 8$. }
\label{fig:N_k=2}
\end{figure}

Figures~\ref{fig:N_k=1} ($N_k=1$ or 1 iteration per time step) and~\ref{fig:N_k=2} ($N_k=2$) show a box and whiskers plot over 10 simulations, with results from a typical simulation overlaid in blue.
In both cases, the Strehl ratio achieves a value $\ge 0.8$ within 150 timeframes.
However, the $N_k=2$ case achieves a slightly higher Strehl ratio with less variation at the cost of approximately twice the computation and latency.

The inset in Figure~\ref{fig:N_k=2} shows the Strehl ratio of a particular simulation versus the time for four different values of $N_k$.
Notice that more iterations per timeframe improves the Strehl ratio but that $N_k=1$ works well and achieves a Strehl ratio of greater than $0.8$.

\begin{figure}
    \centering
    \begin{subfigure}{0.20\textwidth}
         \begin{center}
        \includegraphics[height=1.4in]{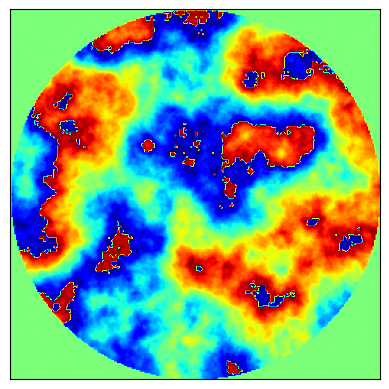}
         \end{center}
         \caption{True phase}
     \end{subfigure}
     \begin{subfigure}{0.25\textwidth}
         \begin{center}
        \includegraphics[height=1.4in]{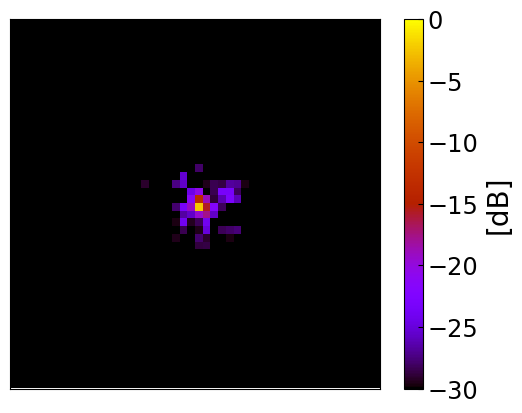}
         \end{center}
         \caption{Residual PSF (zoomed 5.3X) }
     \end{subfigure}
     \vskip 0.1cm
    \begin{subfigure}{0.20\textwidth}
        \begin{center}
        \includegraphics[height=1.4in]{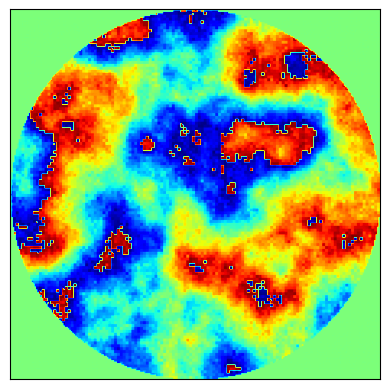}
         \end{center}
         \caption{Est. phase, $N_k=1$ iters per step}
     \end{subfigure}
      \begin{subfigure}{0.25\textwidth}
        \begin{center}
        \includegraphics[height=1.4in]{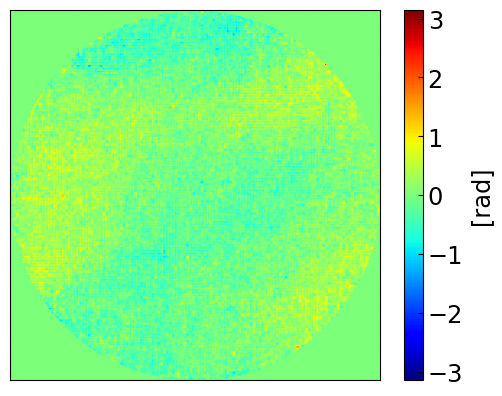}
        \end{center}
         \caption{Residual phase, $N_k=1$ iters per step}
     \end{subfigure}

         
     
    \caption{Phase estimates at time $n=300$, for a particular simulation. Plotted from $-\pi$ to $\pi$, windowed to match the aperture, and with tip and tilt removed. Residual phase is calculated as $\angle e^{j(\phi-\hat{\phi})}$.  Residual PSF is obtained by taking the FFT of the residual phase.  }
\label{fig:3PhasesPlots}
\end{figure}

Figure~\ref{fig:3PhasesPlots} shows the true and estimated phase-errors for time $n=300$ and $N_k=1,2$ iterations per timeframe.
The residual phase plot is computed by subtracting the true and estimated phase, and the residual PSF is computed by taking the inverse FFT of the residual phase.
Both the residual phase and PSF indicate that even with $N_k=1$ the phase is accurately reconstructed.


\begin{figure}
    \centering
         \begin{subfigure}[b]{0.2\textwidth}
         \centering
         \includegraphics[width=1.3\textwidth]{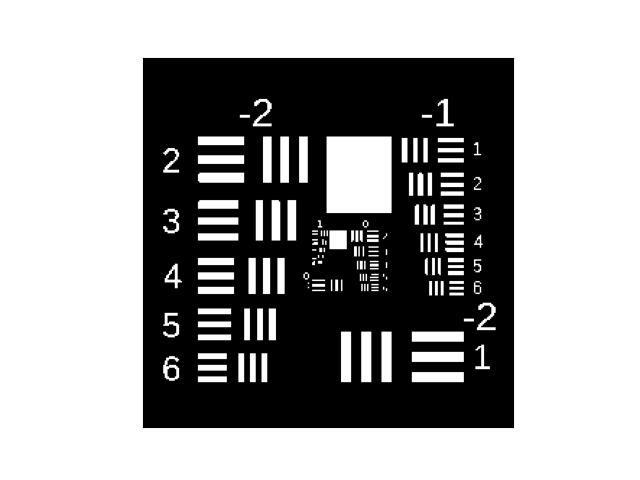}
         \vspace*{-20pt}
         \caption{Ground truth reflectance}
         \label{true r}
     \end{subfigure}
          \begin{subfigure}[b]{0.2\textwidth}
         \centering
         \includegraphics[width=1.3\textwidth]{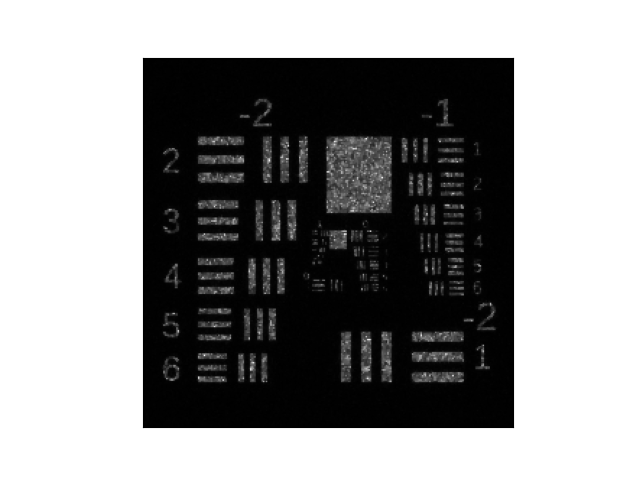}
         \vspace*{-20pt}
         \caption{$N_k=1$ iters/time}
     \end{subfigure}
\caption{Ground truth and reconstructed reflectance estimates at time $n=300$ of a particular simulation. Shown on a scale of $0$ to $1$.}
\label{fig:Reflectances}
\end{figure}

\section{Conclusion}
We described the Dynamic DH-MBIR algorithm for quickly estimating phase-errors due to atmospheric turbulence. Using synthetic DH data generated from frozen flow phase-errors with no boiling, we showed that this algorithm can achieve a Strehl ratio greater than $0.8$ within $150$ timeframes using only $1$ iteration of the EM algorithm per timeframe.
Since the computational cost of each iteration of the Dynamic DH-MBIR algorithm is dominated by two 2D FFTs \cite{SridharKisner2020}, this makes the Dynamic DH-MBIR algorithm practical for real-time implementation.


\bibliographystyle{IEEEtran}
\bibliography{all}

\end{document}